\documentclass[conference]{IEEEtran}
\usepackage[T1]{fontenc}
\usepackage[utf8]{inputenc}
\usepackage{cprotect}
\usepackage{amsmath}
\usepackage{amssymb}
\usepackage{graphicx}
\usepackage[bookmarks=true,bookmarksnumbered=true,bookmarksopen=true,bookmarksopenlevel=1,
 breaklinks=false,pdfborder={0 0 1},backref=false,colorlinks=false]
 {hyperref}
\hypersetup{pdftitle={Your Title},
 pdfauthor={Your Name},
 pdfborderstyle=,pdfpagelayout=OneColumn,pdfnewwindow=true,pdfstartview=XYZ,plainpages=false}

\makeatletter

\usepackage{cprotect}

\usepackage{cite}

\usepackage[font=small,labelfont=small,skip=2pt]{caption}
\ifdefined\showcaptionsetup
 
\fi

\ifdefined\showcaptionsetup
 \PassOptionsToPackage{caption=false}{subfig}
\fi
\usepackage{subfig}
\makeatother

\begin{document}
\title{Physics-Informed Neural Optimization Based Antenna Coding Design for
Pixel Antenna Systems}
\author{\IEEEauthorblockN{Taoning Zhan\IEEEauthorrefmark{1}, Shanpu Shen\IEEEauthorrefmark{2}
and Danny H.K. Tsang\IEEEauthorrefmark{1}} \IEEEauthorblockA{\IEEEauthorrefmark{1}Internet of Thing Thrust, The Hong Kong University
of Science and Technology (Guangzhou), Guangdong, China \\
 \IEEEauthorrefmark{2}State Key Laboratory of Internet of Things
for Smart City, University of Macau, Macau, China\\
 E-mail: \texttt{tzhan@connect.hkust-gz.edu.cn}, \texttt{shanpushen@um.edu.mo}
and \texttt{eetsang@ust.hk}}}
\maketitle
\begin{abstract}
Pixel antennas enable highly radiation pattern reconfigurability to
enhance wireless systems, but its antenna coding design, that is optimizing
the states of switches embedded in pixel antennas, remains an NP-hard
challenge. Conventional approaches for antenna coding design typically
rely on heuristic search algorithms, which suffer from high computational
complexity. To overcome this issue, we propose a novel efficient data-free
optimization algorithm called physics-informed neural optimizer (PINO)
for antenna coding design. By integrating a deep convolutional neural
network prior and a Gumbel-Sigmoid continuous relaxation into a differentiable
physics engine, the proposed algorithm transforms the binary optimization
problem into a continuous differentiable problem, which enables the
antenna coding optimization problem to be efficiently solved via gradient
descent. Simulation results demonstrate that the proposed algorithm
outperforms the heuristic search based algorithms, reducing computational
time while achieving higher average channel gain. 
\end{abstract}


\section{Introduction}

The upcoming sixth generation (6G) communications are envisioned to
support unprecedented requirements in throughput, ultra-low latency,
and massive connectivity \cite{MIMO6G}. To achieve these stringent
requirements, reconfigurable antennas are one of the promising technologies.
Compared with conventional antennas with fixed configurations and
characteristics, reconfigurable antennas offer new degree of freedoms
to adapt to the dynamic and complex propagation environments, thereby
enhancing wireless communication systems.

As a highly reconfigurable antenna technology, pixel antennas are
based on discretizing a continuous radiating surface into a grid of
small sub-wavelength elements, called pixels, interconnected by RF
switches. By controlling the on/off states of these switches, the
topology of pixel antenna can be flexibly reconfigured, thereby enabling
dynamic control of its characteristics such as radiation patterns
\cite{PAintro}, \cite{PAintro2}. A related technology to pixel antennas
is the fluid antenna system (FAS) \cite{FAS}. By optimizing the switch
states, pixel antennas can mimic the position-switching capability
of FAS \cite{11263876} to enhance wireless communication systems.

To fully exploit the potential of pixel antennas, a novel technique
called antenna coding has been recently proposed \cite{AC}, where
the switch states are represented by binary variables called antenna
coder. By optimizing antenna coder, the topology and characteristics
of pixel antennas can be optimized to adapt to channel and enhance
the wireless systems, including \cite{AC,EAC1,EAC2,HLAC,EAC3}. However,
optimizing the binary antenna coding is essentially an NP-hard binary
optimization problem. Existing works \cite{AC,EAC1,EAC2,HLAC} primarily
rely on heuristic search algorithms, such as successive exhaustive
Boolean optimization (SEBO) \cite{SEBO}, but suffer from high computational
complexity. To alleviate this issue, a supervised deep learning approach
using heterogeneous multi-head selection was proposed in \cite{DLAC}.
However, this supervised deep learning approach is date-driven which
still relies on SEBO to generate massive labeled datasets for offline
training. Thus, it remains a challenge to develop an efficient data-free
algorithm for antenna coding design.

To overcome this challenge, we propose a physics-informed neural optimizer
(PINO) based on physics-informed machine learning (PIML) \cite{PIML}.
Operating as an efficient data-free iterative training algorithm,
PINO transforms the intractable binary optimization into a differentiable
continuous problem. By integrating a convolutional neural network
(CNN) and Gumbel-Sigmoid relaxation into a differentiable physics
engine, the proposed algorithm enables concurrent and gradient-based
search of the solution space without requiring any pre-collected datasets.
Through the proposed evaluation metrics and dual convergence criteria,
PINO can efficiently optimize the antenna coding. Simulation results
show that PINO not only significantly reduce the computational complexity
compared to SEBO but also achieves better performance.

\textit{Notations:} Bold lower-case and upper-case letters denote
vectors and matrices, respectively. $\mathbb{C}$ and $\mathbb{R}$
represent the complex sets and real numbers, respectively. $\left|\cdot\right|$
and $\left\Vert \cdot\right\Vert $ denote the modulus and the $\ell_{2}$-norm
of a vector, respectively. $(\cdot)^{T}$, $(\cdot)^{-1}$ and $\left[\cdot\right]_{i,j}$
denote the transpose, inverse and $(i,j)$th element of a matrix,
respectively. $\Re\{\cdot\}$ and $\Im\{\cdot\}$ represent the real
parts and imaginary parts, respectively. $\mathcal{\mathcal{C}N}(0,1)$
represent the circularly symmetric complex Gaussian (CSCG) distribution
with zero mean and identity covariance value.

\section{Antenna Coding Based on Pixel Antenna}\label{sec:PM}

In this section, we introduce the physics model of pixel antenna and
the formulation of antenna coding design. 
\begin{figure}[t]
\begin{centering}
\subfloat[]{\begin{centering}
\includegraphics[bb=-1bp -40bp 990bp 1169bp,clip,scale=0.09]{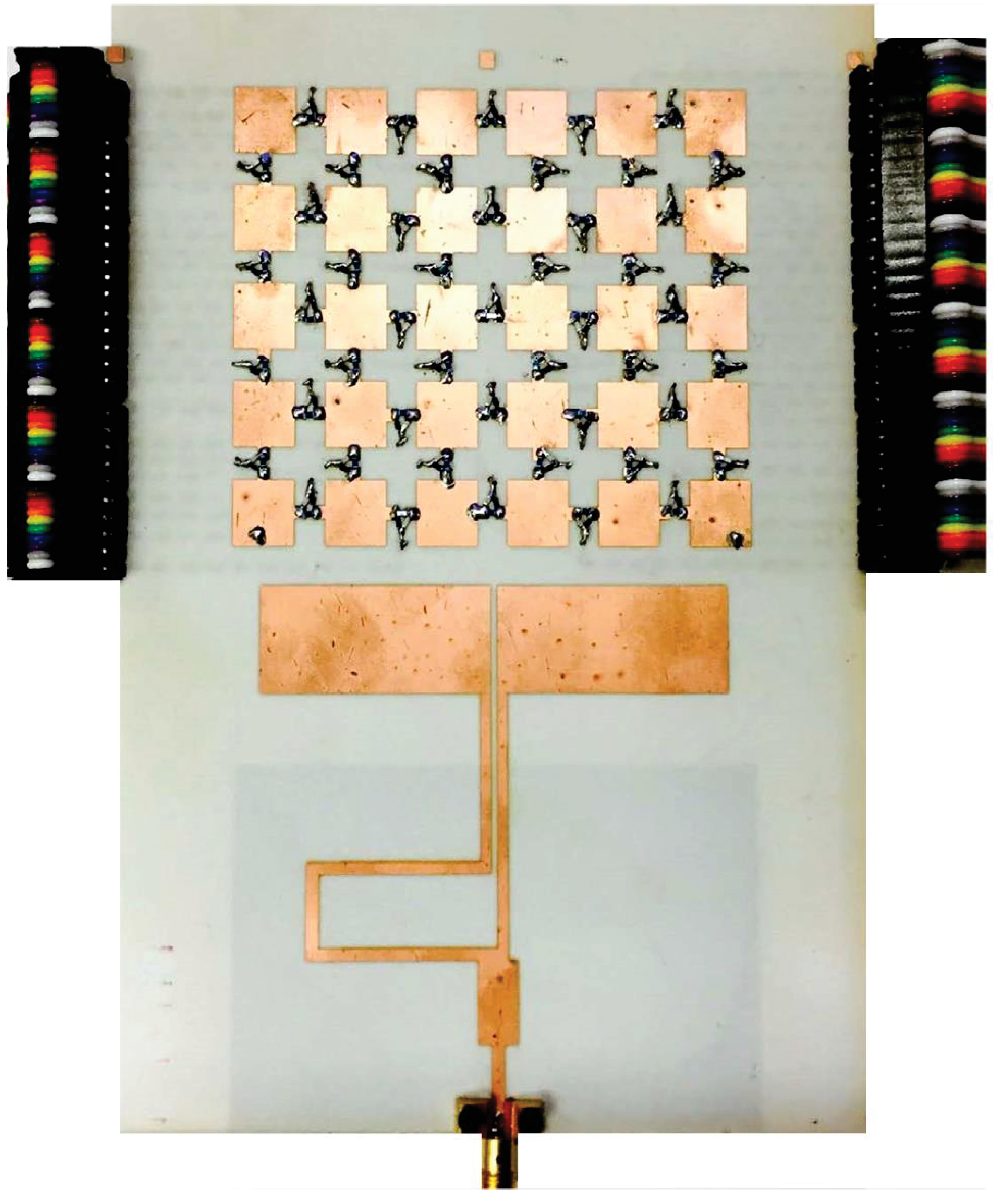} 
\par\end{centering}
}\subfloat[]{\begin{centering}
\includegraphics[scale=0.065]{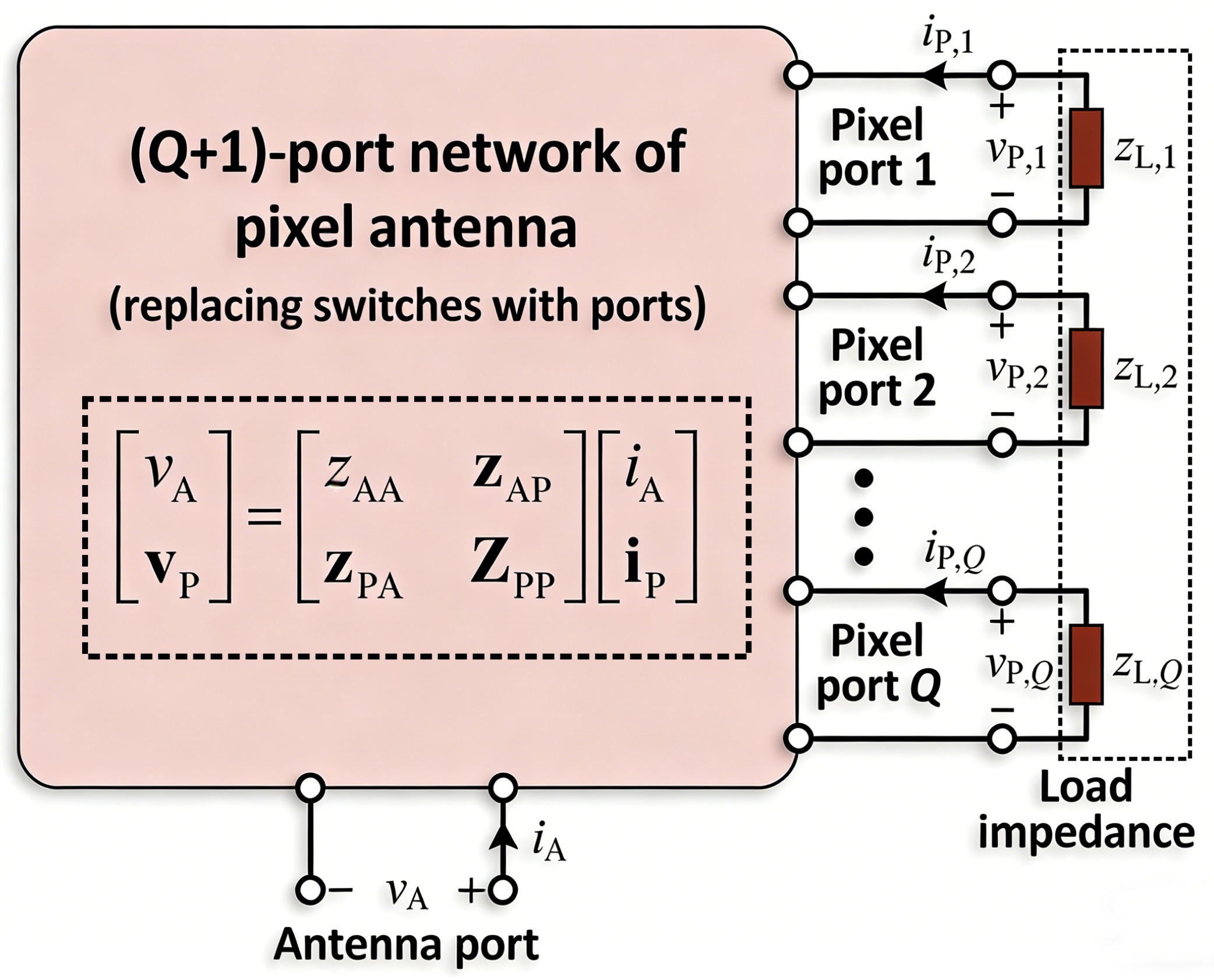} 
\par\end{centering}
}
\par\end{centering}
\caption{(a) An example of pixel antenna prototype in \cite{PAprototype}.
(b) Multi-port circuit network model for pixel antenna.}
\end{figure}

\subsection{Physics Model of Pixel Antenna}

As shown in Fig. 1(a), a pixel antenna contains a grid of pixels embedded
with RF switches. Consider a pixel antenna with $Q$ switches, we
can model it as a $(Q+1)$-port circuit network including one antenna
port and $Q$ pixel ports replacing the $Q$ switches, as shown in
Fig. 1(b), which is characterized by its impedance matrix $\mathbf{Z}\in\mathbb{C}^{(Q+1)\times(Q+1)}$,
given by 
\begin{equation}
\mathbf{Z}=\left[\begin{array}{ll}
z_{\textrm{AA}} & \mathbf{z}_{\textrm{AP}}\\
\mathbf{z}_{\textrm{PA}} & \mathbf{Z}_{\textrm{PP}}
\end{array}\right],\label{eq:MPT-1}
\end{equation}
where $z_{\textrm{AA}}\in\mathbb{C}$ is the self impedance for the
antenna port, $\mathbf{Z}_{\textrm{PP}}\in\mathbb{C}^{Q\times Q}$
is the impedance matrix for the $Q$ pixel ports, and $\mathbf{z}_{\textrm{AP}}\in\mathbb{C}^{1\times Q}$
with its transpose $\mathbf{z}_{\textrm{PA}}\in\mathbb{C}^{Q\times1}$
are the trans-impedance between the antenna port and the pixel ports.
Thus, we can relate the voltage and current at all ports by 
\begin{equation}
\left[\begin{array}{c}
v_{\textrm{A}}\\
\mathbf{v}_{\textrm{P}}
\end{array}\right]=\left[\begin{array}{ll}
z_{\textrm{AA}} & \mathbf{z}_{\textrm{AP}}\\
\mathbf{z}_{\textrm{PA}} & \mathbf{Z}_{\textrm{PP}}
\end{array}\right]\left[\begin{array}{c}
i_{\textrm{A}}\\
\mathbf{i}_{\textrm{P}}
\end{array}\right],\label{eq:MPT}
\end{equation}
where $v_{\textrm{A}}\in\mathbb{C}$ and $i_{\textrm{A}}\in\mathbb{C}$
are the voltage and current at the antenna port, respectively, and
$\mathbf{v}_{\textrm{P}}=[v_{\textrm{P},1},\ldots,v_{\textrm{P},Q}]^{T}\in\mathbb{C}^{Q\times1}$
and $\mathbf{i}_{\textrm{P}}=[i_{\textrm{P},1},\ldots,i_{\textrm{P},Q}]^{T}\in\mathbb{C}^{Q\times1}$
are the voltages and currents at the $Q$ pixel ports, respectively.

Each pixel port is in series with a switch, which has two states,
on and off. Thus, we can model the $q$th switch by load impedance
$z_{\textrm{L},q}$, which is either short- or open-circuit, and use
a binary variable $b_{q}\in\left\{ \textrm{0,1}\right\} $ to represent
the state of the $q$th switch, $\forall q\in\mathcal{Q}\triangleq\{1,2,...,Q\}$,
which is given by 
\begin{equation}
z_{\textrm{L},q}=\begin{cases}
0, & b_{q}=0,\:\textrm{i.e. short-circuit,}\\
\infty, & b_{q}=1,\:\textrm{i.e. open-circuit}.
\end{cases}
\end{equation}
We define the vector $\mathbf{b}=[b_{1},\ldots,b_{Q}]^{T}\in\left\{ \textrm{0,1}\right\} ^{Q\times1}$
as the antenna coder, which describes the states of all switches for
the pixel antenna. We collect $z_{\textrm{L},q}$ $\forall q$ into
a diagonal matrix $\mathbf{Z}_{\textrm{L}}\left(\mathbf{b}\right)=\textrm{diag}(z_{\textrm{L},1},\ldots,z_{\textrm{L},Q})\in\mathbb{C}^{Q\times Q}$
such that the voltage and current at pixel ports satisfy $\mathbf{v}_{\textrm{P}}=-\mathbf{Z}_{\textrm{L}}\left(\mathbf{b}\right)\mathbf{i}_{\textrm{P}}$.
Substituting this into (\ref{eq:MPT}), the current at pixel ports
can be expressed as a function of the antenna coder $\mathbf{b},$
i.e. 
\begin{equation}
\mathbf{i}_{\textrm{P}}\left(\mathbf{b}\right)=-\left(\mathbf{Z}_{\textrm{PP}}+\mathbf{Z}_{\textrm{L}}\left(\mathbf{b}\right)\right)^{-1}\mathbf{z}_{\textrm{PA}}i_{\textrm{A}}.
\end{equation}

The radiation pattern of pixel antenna is a superposition of the patterns
from all ports, weighted by the currents. Let $\mathbf{E}_{\textrm{oc}}=[\mathbf{e}_{\textrm{A}},\mathbf{e}_{\textrm{P},1},\ldots,\mathbf{e}_{\textrm{P},Q}]\in\mathbb{C}^{2K\times(Q+1)}$
denote the open-circuit radiation pattern matrix, where $\mathbf{e}_{\textrm{A}}\in\mathbb{C}^{2K\times1}$
and $\mathbf{e}_{\textrm{P},q}\in\mathbb{C}^{2K\times1}$ represent
the radiation patterns of the antenna port and pixel ports (including
both $\theta$ and $\phi$ polarization components over $K$ spatial
angles) when a unit current excites the corresponding port while all
other ports are open-circuit. As such, the radiation pattern of pixel
antenna $\mathbf{e}\left(\mathbf{b}\right)\in\mathbb{C}^{2K\times1}$
can be expressed as 
\begin{equation}
\mathbf{e}\left(\mathbf{b}\right)=\mathbf{E}_{\textrm{oc}}\mathbf{i}\left(\mathbf{b}\right),\label{eq:EI}
\end{equation}
where $\mathbf{i}\left(\mathbf{b}\right)=[i_{\textrm{A}};\mathbf{i}_{\mathrm{P}}\left(\mathbf{b}\right)]\in\mathbb{C}^{(Q+1)\times1}$
collects current at all ports. It can found from (\ref{eq:EI}) that
by optimizing antenna coder $\mathbf{b}$, we can flexibly reconfigure
the radiation pattern $\mathbf{e}\left(\mathbf{b}\right)$. 
\begin{figure}[t]
\begin{centering}
\includegraphics[scale=0.4]{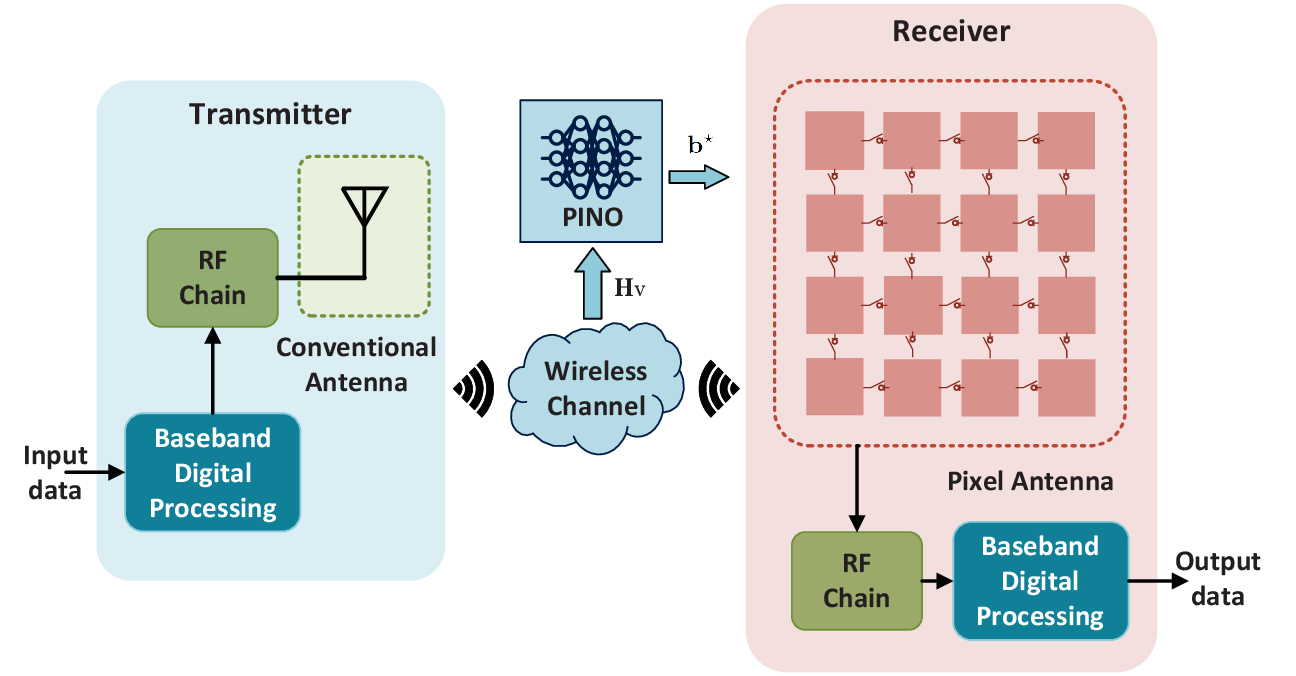} 
\par\end{centering}
\caption{Diagram of SISO pixel antenna system integrated with the PINO.}
\end{figure}

\cprotect

\cprotect\subsection{Antenna Coding Design Formulation 
\begin{figure*}[t]
\protect
\protect\begin{centering}
\protect\protect\includegraphics[scale=0.88]{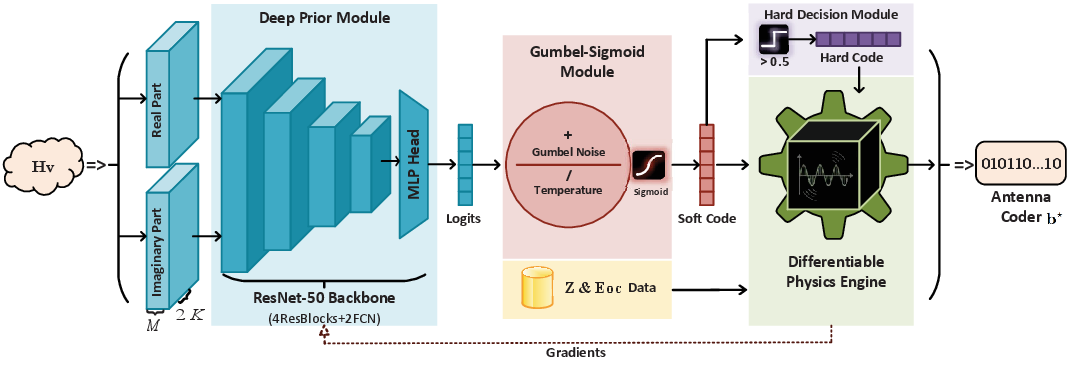}\protect \protect
\par\end{centering}
\protect\protect\caption{The architecture of PINO. }\label{fig:pinn}
\end{figure*}
}

In this work, we focus on a pixel antenna empowered single-input single-output
(SISO) system where the transmitter uses a conventional antenna and
the receiver uses a pixel antenna, as illustrated in Fig. 2. We use
the beamspace channel representation to model the channel of pixel
antenna empowered SISO system \cite{AC}, written as 
\begin{equation}
h\left(\mathbf{b}_{\textrm{R}}\right)=\mathbf{e}_{\textrm{R}}^{T}\left(\mathbf{b}_{\textrm{R}}\right)\mathbf{H}_{\textrm{V}}\mathbf{e}_{\textrm{T}},
\end{equation}
where $\mathbf{e}_{\textrm{T}}\in\mathbb{C}^{2K\times1}$ is the normalized
radiation pattern of the transmit antenna, $\mathbf{e}_{\textrm{R}}\left(\mathbf{b}_{\textrm{R}}\right)\in\mathbb{C}^{2K\times1}$
is the normalized radiation pattern of the receive pixel antenna,
which is coded by $\mathbf{b}_{\textrm{R}}$, satisfying $\left\Vert \mathbf{e}_{\textrm{R}}\left(\mathbf{b}_{\textrm{R}}\right)\right\Vert =\left\Vert \mathbf{e}_{\textrm{T}}\right\Vert =1$,
and $\mathbf{H}_{\textrm{V}}\in\mathbb{C}^{2K\times2K}$is the virtual
channel matrix, with each entry being the channel gain from an angle
of departure (AoD) to an angle of arrival (AoA) over $K$ spatial
angles and two polarizations.

Aiming at maximizing the channel gain, the antenna coding design problem
can be formulated as 
\begin{align}
\underset{\mathbf{b}_{\textrm{R}}}{\text{max}}\ \  & \left|\mathbf{e}_{\textrm{R}}^{T}\left(\mathbf{b}_{\textrm{R}}\right)\mathbf{H}_{\textrm{V}}\mathbf{e}_{\textrm{T}}\right|^{2}\label{eq:SISO}\\
\text{s.t.}\ \ \  & \left[\mathbf{b}_{\textrm{R}}\right]_{q}\in\left\{ 0,1\right\} ,\forall q\in\mathcal{Q},\label{eq:SISO1}
\end{align}
which is an NP-hard binary optimization problem.

\section{Physics-Informed Neural Optimization}

In this section, we propose a physics-informed neural optimizer to
solve the antenna coding design problem (\ref{eq:SISO})-(\ref{eq:SISO1}).
We first show the architecture of PINO and then explain the overall
online training process.

\subsection{Architecture of PINO}\label{subsec:Architecture-of-PINS}

To enable gradient-based optimization, we construct a differentiable
forward-pass architecture comprising five modules, as shown in Fig.
\ref{fig:pinn}.

\textit{1) Tensor Initialization Module:} The proposed PINO architecture
begins with input preprocessing. To leverage real-valued convolutional
neural networks, the instantaneous complex virtual channel matrix
$\mathbf{H}_{\textrm{V}}\in\mathbb{C}^{2K\times2K}$ is first decoupled
into a two-channel real-valued spatial feature tensor $\mathbf{x}$
\begin{equation}
\mathbf{x}=\left\{ \Re\{\mathbf{H}_{\mathrm{V}}\},\Im\{\mathbf{H}_{\mathrm{V}}\}\right\} \in\mathbb{R}^{2\times2K\times2K}.\label{eq:Tensor}
\end{equation}
To avoid local optimal solution and facilitate a multi-start concurrent
search, we duplicate the input feature $\ensuremath{\mathbf{x}}$
to construct a batched input tensor $\mathbf{X}=\{\mathbf{x}^{(1)},\mathbf{x}^{(2)}\ldots,\mathbf{x}^{(M)}\}\in\mathbb{R}^{M\times2\times2K\times2K},$
where $\mathbf{x}^{(m)}=\mathbf{x}$, $\forall m\in\{1,2,...,M\},$
thereby instantiating $\ensuremath{M}$ independent concurrent search
probes.

\textit{2) Deep Prior Module (ResNet-50+MLP)}: The batched tensor
$\ensuremath{\mathbf{X}}$ is fed into a modified ResNet-50 backbone
\cite{7780459}, where the first convolutional layer is adapted to
accept the 2-channel input. Acting as a structural deep prior, the
network extracts high-dimensional spatial correlation features of
the channel through its residual blocks, and then passed through a
multi-layer perceptron (MLP) projection head, compressing them to
map exactly to the $Q$ pixel ports. The final output is an unnormalized
probability matrix (logits), which is given by 
\begin{equation}
\mathbf{W}=f_{\boldsymbol{\rho}}(\mathbf{X})\in\mathbb{R}^{Q\times M},
\end{equation}
where $\ensuremath{f_{\boldsymbol{\rho}}(\cdot)}$ and $\boldsymbol{\rho}$
are the mapping function and trainable weights of the neural network,
respectively, and $\mathbf{W}=[\mathbf{w}^{(1)},\dots,\mathbf{w}^{(M)}]$
with $\mathbf{w}^{(m)}=[w_{1}^{(m)},\dots,w_{Q}^{(m)}]^{T}\in\mathbb{R}^{Q\times1}$
collecting all the logits for the $m$th probe.

\textsl{3) }\textit{Gumbel-Sigmoid Relaxation} \textit{Module:} Since
the antenna coder $\mathbf{b}$ is binary, backpropagation of the
deep prior module cannot be directly applied. Therefore, we introduce
the Gumbel-Sigmoid continuous relaxation mechanism. Specifically,
for the $q$th port of the $m$th probe, we use an independent random
variable $u_{q}^{(m)}\sim\text{Uniform}(0,1)$ to generate the logistic
Gumble noise 
\begin{equation}
g_{q}^{(m)}=\ln(u_{q}^{(m)})-\ln(1-u_{q}^{(m)}).
\end{equation}
This noise is injected element-wise into the corresponding logit $w_{q}^{(m)}$
to break symmetry and encourage stochastic exploration in the solution
space. Activated by a Sigmoid function $\ensuremath{\sigma(\cdot)}$,
the normalized probability for the $q$th port of the $m$th probe
$\hat{b}_{q}^{(m)}\in(0,1)$ can be obtained by 
\begin{equation}
\hat{b}_{q}^{(m)}=\sigma\left(\frac{w_{q}^{(m)}+g_{q}^{(m)}}{\tau}\right)=\frac{1}{1+e^{-\frac{w_{q}^{(m)}+g_{q}^{(m)}}{\tau}}},
\end{equation}
where $\tau\in\mathbb{R}^{+}$ is a global temperature annealing parameter,
controlling the steepness of the Sigmoid function. To ensure this
continuous approximation eventually approaching the binary extremities,
i.e. $\hat{b}_{q}^{(m)}\rightarrow\{0,1\}$, $\tau$ follows an exponential
temperature decay schedule over training epochs $\ensuremath{t}$,
i.e. 
\begin{equation}
\tau^{(t)}=\max(\tau_{\min},\tau_{0}\gamma^{t}),\label{eq:temperature}
\end{equation}
where $\tau_{0}$ is the initial temperature, $\gamma\in(0,1)$ is
the decay rate, and $\tau_{\min}$ is the lower bound. For each concurrent
search probe, we group these normalized probabilities and defined
as the soft code vector $\hat{\mathbf{b}}^{(m)}=[\hat{b}_{1}^{(m)},\dots,\hat{b}_{Q}^{(m)}]^{T}\in(0,1)^{Q\times1}$,
which represents a transition state during training. This relaxation
allows the discrete on/off switch states to be smoothly approximated,
building a differentiable mathematical bridge.

\textsl{4) }\textit{Hard Decision Module:} To recover the true physical
switch states during evaluation stage, we introduce a hard decision
mechanism. Specifically, for each soft code of the $m$th probe $\hat{\mathbf{b}}^{(m)}$,
we map this continuous soft code into a discrete hard code 
\begin{equation}
\mathbf{b}^{(m)}=\mathbb{I}\left(\hat{\mathbf{b}}^{(m)}>0.5\right)\enspace\forall m,
\end{equation}
where $\mathbb{I}(\cdot$) is the indicator function. Thus, we have
$\mathbf{b}^{(m)}=[b_{1}^{(m)},\ldots,b_{Q}^{(m)}]^{T}\in\left\{ \textrm{0,1}\right\} ^{Q\times1}$,
i.e. the antenna coder.

\textsl{5) }\textit{Differentiable Physics Engine:} This module calculates
the objective functions (\ref{eq:SISO}) with the obtained soft and
hard code based on the physics model of pixel antenna introduced in
Section \ref{sec:PM}. We define the evaluation metrics as follows:

Soft gain: With the soft code $\hat{\mathbf{b}}^{(m)}\in(0,1)^{Q\times1}$,
the soft gain for gradient guidance is calculated by $G_{\mathrm{soft}}^{(m)}=\mathcal{F}_{\text{phy}}\left(\hat{\mathbf{b}}^{(m)},\mathbf{H}_{\text{V}}\right)=\left|\mathbf{e}_{\textrm{R}}^{T}\left(\hat{\mathbf{b}}^{(m)}\right)\mathbf{H}_{\textrm{V}}\mathbf{e}_{\textrm{T}}\right|^{2},$
where $\mathcal{F}_{\text{phy}}$ denotes the differentiable function
and $\hat{\mathbf{b}}^{(m)}$ is mapped to continuous load impedance
via a logarithmic interpolation between the discrete on and off switch
states to ensure stable gradient backpropagation.

Hard gain: With the corresponding hard code $\mathbf{b}^{(m)}\in\left\{ \textrm{0,1}\right\} ^{Q\times1}$,
the hard gain representing the true physical performance is evaluated
as $G_{\mathrm{hard}}^{(m)}=\left|\mathbf{e}_{\textrm{R}}^{T}\left(\mathbf{b}^{(m)}\right)\mathbf{H}_{\textrm{V}}\mathbf{e}_{\textrm{T}}\right|^{2}.$

Loss function: The objective of optimizer is defined as 
\begin{equation}
\mathcal{L}(\boldsymbol{\rho})=-\frac{1}{M}\sum_{m=1}^{M}\mathcal{F}_{\text{phy}}(\hat{\mathbf{b}}^{(m)},\mathbf{H}_{\mathrm{V}}).
\end{equation}

By minimizing the loss function $\mathcal{L}(\boldsymbol{\rho})$
through iterative training, the network parameters \textbf{$\boldsymbol{\rho}$}
are continuously updated via gradient descent, guiding the concurrent
probes to explore the continuous solution space, and converge toward
the optimal antenna coder that maximizes the averaged beamspace channel
gain over all the concurrent probes.

\subsection{Training Process}\label{subsec:Optimization-and-Training}

The training of PINO is an iterative process executed directly on
an input channel sample. Based on the architecture defined above,
the online training logic proceeds through the following four steps.

Step 1: Construct the batched tensor $\ensuremath{\mathbf{X}}$ from
the input channel matrix $\ensuremath{\mathbf{H}_{\mathrm{V}}}$,
and randomly initialize the neural network weights $\boldsymbol{\rho}$.
Define the maximum number of epochs $\ensuremath{T_{\mathrm{max}}}$,
initial temperature $\ensuremath{\tau_{0}}$, and convergence patience
window $\ensuremath{P}$.

Step 2: At the current training epoch $\ensuremath{t}\left(\ensuremath{t}\leq\ensuremath{T_{\mathrm{max}}}\right)$,
update the temperature $\tau^{(t)}$ based on the exponential annealing
schedule (\ref{eq:temperature}). For each probes, the network performs
a forward pass to output logits $\mathbf{w}^{(m,t)}$, generating
the current soft codes $\ensuremath{\hat{\mathbf{b}}^{(m,t)}}$ combined
with injected noise. The physics engine then evaluates the soft gain
for each probe $G_{\mathrm{soft}}^{(m,t)}$, yielding the average
soft gain $\ensuremath{\bar{G}_{\mathrm{soft}}^{(t)}=-\frac{1}{M}\sum_{m=1}^{M}G_{\mathrm{soft}}^{(m,t)}}$
and the total loss $\ensuremath{\mathcal{L}^{(t)}(\boldsymbol{\rho})}$
of current training epoch.

Step 3: Compute gradients $\ensuremath{\nabla_{\boldsymbol{\rho}}\mathcal{L}^{(t)}}$
via Autograd through\textit{ }physics engine $\mathcal{F}_{\text{phy}}$.
These gradients are backpropagated to update the network weights $\boldsymbol{\rho}$
via the AdamW optimizer, driving the collective search of the concurrent
probes on the continuous manifold.

Step 4: After updating the parameters, the gradient computation graph
is temporarily disabled for evaluation. The hard decision is applied
to extract the hard codes $\ensuremath{\mathbf{b}^{(m,t)}}$, which
are fed into the physics engine to compute the current average hard
gain $\ensuremath{\bar{G}_{\mathrm{hard}}^{(t)}=-\frac{1}{M}\sum_{m=1}^{M}G_{\mathrm{hard}}^{(m,t)}}$.
The system then evaluates the steady-state dual convergence criteria
\begin{equation}
\left\{ \begin{aligned}\Delta_{\text{loss}}^{(t)}= & \left|\mathcal{L}^{(t)}(\boldsymbol{\rho})-\mathcal{L}^{(t-P)}(\boldsymbol{\rho})\right|<\epsilon_{\mathcal{L}},\\
\Delta_{\text{gap}}^{(t)}= & \left|\bar{G}_{\text{soft}}^{(t)}-\bar{G}_{\text{hard}}^{(t)}\right|<\epsilon_{\text{gap}},
\end{aligned}
\right.
\end{equation}
where $\epsilon_{\mathcal{L}}$ and $\epsilon_{\text{gap}}$ are the
tolerance threshold of the loss function and the performance gap between
the soft and hard gain, respectively. Once both criteria are satisfied
at epoch $T_{\text{conv}}$, the network converges and the training
terminates. The optimizer then outputs the antenna coder by selecting
the probe yielding the maximum hard gain among the converged ensemble
\begin{equation}
\mathbf{b}^{\star}=\textrm{arg}\max_{\mathbf{b}^{(m,T_{\text{conv}})}}G_{\mathrm{hard}}^{(m,T_{\text{conv}})}.
\end{equation}

The overall complexity of PINO algorithm is given by $\mathcal{O}\left(MT_{\text{conv}}^{(M)}\right)$,
where $T_{\text{conv}}^{(M)}$ denotes the required convergence epochs
for $M$ concurrent probes. Since both variables are independent of
the pixel port number $\ensuremath{Q}$, PINO avoids the high complexity
of heuristic search algorithms such as SEBO.

\section{Performance Evaluation}

In this section, we evaluate the performance of the antenna coding
design based on the proposed PINO.

\subsection{Simulation Settings}

We consider a rich scattering propagation environment with Rayleigh
fading, where each entry of the beamspace channel $[\mathbf{H}_{\text{V}}]_{i,j}\,\forall i,j$
are independent and identically distributed (i.i.d.) random variables
following $\mathcal{\mathcal{C}N}(0,1)$. The number of spatial angles
is set as $K=72$. For the SISO system, we consider an isotropic antenna
at the transmitter and a pixel antenna at the receiver. Following
the design in \cite{AC}, the received pixel antenna operates at $2.4$
GHz with a physical aperture $0.5\lambda\times0.5\lambda$ and $Q=39$
pixel ports, where $\lambda=125$ mm denotes the wavelength. We use
the CST studio suite to obtain the the impedance matrix $\mathbf{Z}\in\mathbb{C}^{(Q+1)\times(Q+1)}$
and the open-circuit radiation pattern matrix $\mathbf{E}_{\textrm{oc}}\in\mathbb{C}^{2K\times(Q+1)}$
of this $(Q+1)$ port network. For the PINO, we set learning rate
of the AdamW optimizer as $5\mathrm{e}^{-4},$ the maximum number
of training epoch $\ensuremath{T_{\mathrm{max}}}=500$, initial temperature
$\ensuremath{\tau_{0}}=1.0$ with the decay rate $\gamma=0.998$,
and the convergence patience window $\ensuremath{P}=10$ with the
tolerance threshold $\epsilon_{\mathcal{L}}=\epsilon_{\text{gap}}=1\mathrm{e}^{-3}$.

\subsection{System Performance}

In Fig. 4, we first evaluate the convergence behavior of the proposed
PINO over a fixed channel sample. As shown in Fig. 4(a), the maximum
soft/ hard gains initially fluctuate at high levels due to the Gumbel
noise. As the temperature decays, the shared-weight CNN implicitly
averages the spatial gradients, guiding all probes toward the probe
that achieve the maximum gain. Finally, the average and maximum soft/hard
gains perfectly align around epoch $130$, successfully triggering
the dual convergence criteria. Fig. 1(b) investigates the impact of
$\ensuremath{M}$ on the optimization performance. Increasing $\ensuremath{M}$
from 8 to 64 significantly enhances the converged channel gain by
expanding the concurrent search capability. However, this improvement
exhibits a ceiling effect, plateauing after $\ensuremath{M=64}$.
Concurrently, a larger $\ensuremath{M}$ drastically reduces the number
of epochs required for convergence, as a larger batch size provides
more stable and accurate gradient estimations for the neural network.
\begin{figure}[t]
\begin{centering}
\subfloat[]{\begin{centering}
\includegraphics[bb=37bp 0bp 275bp 237bp,width=3.3cm]{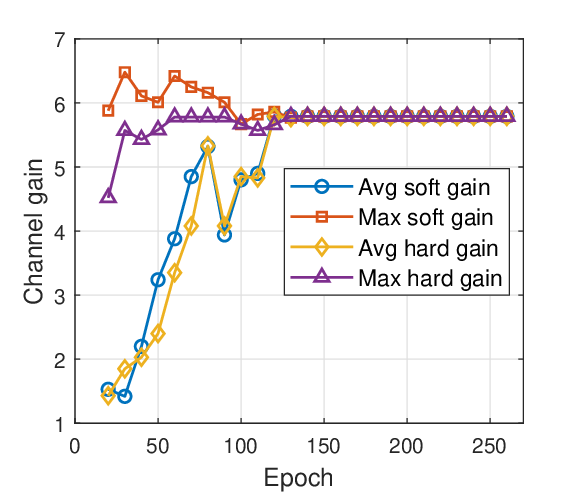} 
\par\end{centering}
\centering{}

}\subfloat[]{\centering{}\includegraphics[bb=38bp 0bp 293bp 241bp,width=3.4cm]{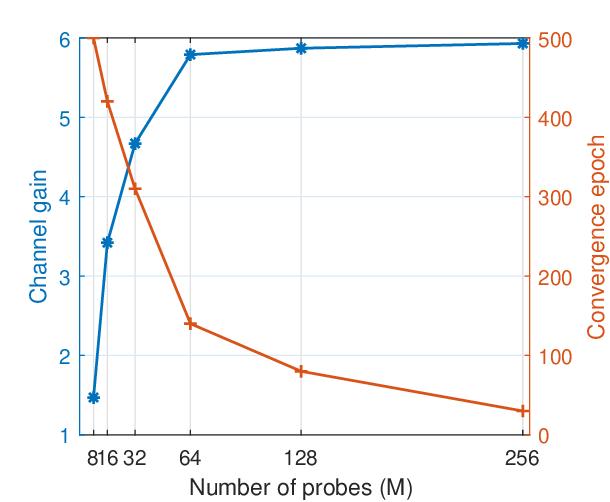}

}
\par\end{centering}
\caption{Performance of the proposed PINO given a fixed channel sample. (a)
Convergence trajectories with the number of probes $M=64$. (b) Impact
of $\ensuremath{M}$ on the channel gain and required convergence
epochs.}
\end{figure}

\begin{figure}[!t]
\begin{centering}
\subfloat[]{\begin{centering}
\includegraphics[bb=15bp 0bp 246bp 236bp,width=3.4cm]{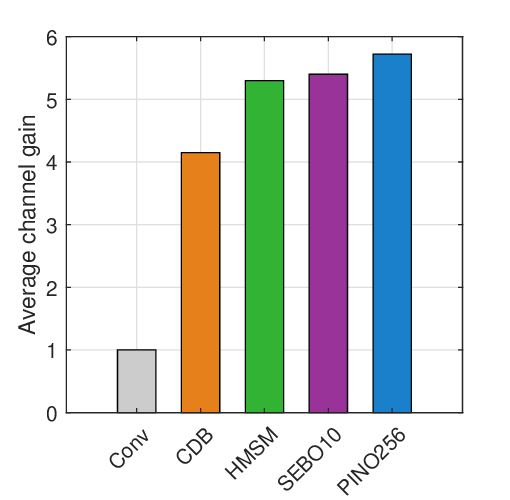} 
\par\end{centering}
}\subfloat[]{\begin{centering}
\includegraphics[bb=30bp 0bp 270bp 258bp,width=3.3cm]{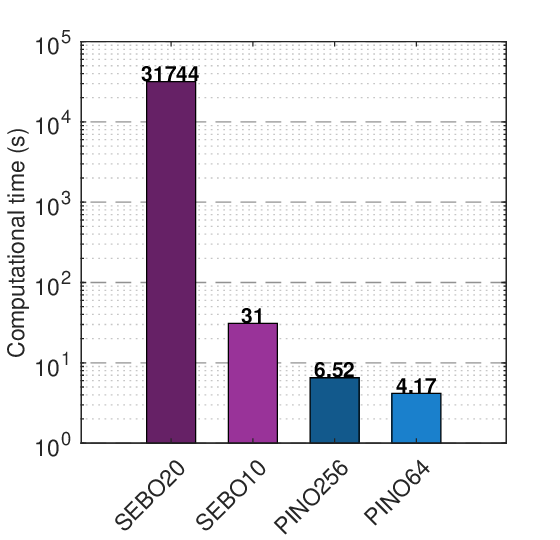} 
\par\end{centering}
}
\par\end{centering}
\caption{Performance comparison among different algorithms. (a) Average channel
gain. (b) Computational time.}
\end{figure}

In Fig. 5, we compare the performance of PINO with existing algorithms.
As shown in Fig. 5(a), PINO with $M=256$ achieves the highest average
channel gain. It outperforms the conventional SISO system using fixed
antenna (Conv), offline codebook designs with the codebook size of
1024 (CDB)\cite{AC}, the supervised learning method (HMSM) \cite{DLAC},
and the SEBO algorithm with a block size $\ensuremath{J=10}$ \cite{SEBO}.
PINO outperforms the SEBO-based algorithms because its continuous
gradient descent is navigable, which avoids the poor local optimal
solution that is often trapped in searching algorithms. Fig. 5(b)
compares the computational time of PINO with SEBO, where both algorithms
are data-free optimization. CDB and HMSM are excluded because they
are date-driving requiring massive offline training. For SEBO, increasing
the search block size from $\ensuremath{J=10}$ to $\ensuremath{J=20}$
causes the computational time to explode exponentially. In contrast,
the computational time of PINO with $M=256$ is only 56\% longer than
that of PINO with $M=64$, benefiting from the gradient-based neural
network structure. The results demonstrate that PINO achieves superior
channel gain while significantly reducing the complexity, making it
practical for antenna coding design.

\section{Conclusion}

In this paper, we propose a physics-informed neural optimizer to address
the NP-hard antenna coding design problem for SISO pixel antenna systems.
By integrating a convolutional neural network and Gumbel-Sigmoid continuous
relaxation into a differentiable physics engine, PINO transforms the
binary problem into a continuous problem, which enables highly parallelized,
data-free optimization solved by gradient descent method. Simulation
results demonstrate that PINO achieves a higher average channel gain
than heuristic search algorithms while reducing computational complexity.
This work establishes PINO as an efficient and promising solution
for practical antenna coding design.

\section*{Acknowledgment}

This work was supported in part by Guangdong Provincial Project under
Grant 2021JC02X149, in part by Guangzhou Municipal Science and Technology
Project under Grant 2023A03J0011, in part by Guangzhou Municipal Key
Laboratory on Future Networked Systems (024A03 J0623), in part by
Guangdong Provincial Key Laboratory of Integrated Communications,
Sensing and Computation for Ubiquitous Internet of Things (No.2023B1212010007),
and in part by the Science and Technology Development Fund, Macau
SAR (File/Project no. 001/2024/SKL).

 \bibliographystyle{IEEEtran}
\bibliography{Ref}

@ARTICLE{11263876,
  author={Wong, Kai-Kit and Wang, Chao and Shen, Shanpu and Chae, Chan-Byoung and Murch, Ross},
  journal={IEEE Wireless Commun}, 
  title={Reconfigurable Pixel Antennas Meet Fluid Antenna Systems: A Paradigm Shift to Electromagnetic Signal and Information Processing}, 
  year={2025},
  volume={},
  number={},
  pages={1-8},
  doi={10.1109/MWC.2025.3625130}
}

@INPROCEEDINGS{HLAC,
  author={Li, Hongyu and Shen, Shanpu},
  booktitle={2025 IEEE 26th International Workshop on Signal Processing and Artificial Intelligence for Wireless Communications (SPAWC)}, 
  title={Antenna Coding Design Based on Pixel Antennas for Multi-User {MISO} Systems}, 
  year={2025},
  volume={},
  number={},
  pages={1-5},
  doi={10.1109/SPAWC66079.2025.11143412}
}

@ARTICLE{AC,
  author={Shen, Shanpu and Wong, Kai-Kit and Murch, Ross},
  journal={IEEE Trans.Commun}, 
  title={Antenna Coding Empowered by Pixel Antennas}, 
  year={2026},
  volume={74},
  number={},
  pages={446-460},
  doi={10.1109/TCOMM.2025.3621089}
}

@INPROCEEDINGS{7780459,
  author={He, Kaiming and Zhang, Xiangyu and Ren, Shaoqing and Sun, Jian},
  booktitle={2016 IEEE Conference on Computer Vision and Pattern Recognition (CVPR)}, 
  title={Deep Residual Learning for Image Recognition}, 
  year={2016},
  volume={},
  number={},
  pages={770-778},
  doi={10.1109/CVPR.2016.90}
}

@article{DLAC,
  author  = {Binzhou Zuo and Shanpu Shen and Hongyu Li},
  title   = {Antenna Coding Optimization for Pixel Antenna Empowered Wireless Communication Using Deep Learning with Heterogeneous Multi-Head Selection},
  journal = {arXiv:2602.23831},
  year    = {2026}
}

@article{EAC1,
  author  = {Tianrui Qiao and Shanpu Shen and Yijun Chen and Ross Murch},
  title   = {Optimizing Antenna Coding for Pixel Antenna Empowered {SISO-OFDM} Systems},
  journal = {arXiv:2603.17658},
  year    = {2026}
}

@article{EAC2,
  author  = {Yijun Chen and Shanpu Shen and Tianrui Qiao and Hongyu Li and Kai-Kit Wong and Ross Murch},
  title   = {Antenna Coding Optimization for Pixel Antenna Empowered {MIMO} Wireless Power Transfer},
  journal = {arXiv:2601.07324},
  year    = {2026}
}

@ARTICLE{EAC3,
  author={Han, Zixiang and Shen, Shanpu and Murch, Ross},
  journal={IEEE Journal of Selected Topics in Signal Processing}, 
  title={Exploiting Spatial Multiplexing Based on Pixel Antennas: An Antenna Coding Approach}, 
  year={2026},
  volume={},
  number={},
  pages={1-14},
  doi={10.1109/JSTSP.2026.3657637}
}

@ARTICLE{SEBO,
  author={Shen, Shanpu and Sun, Ying and Song, Sichao and Palomar, Daniel P. and Murch, Ross D.},
  journal={IEEE Trans. Antennas Propag}, 
  title={Successive Boolean Optimization of Planar Pixel Antennas}, 
  year={2017},
  volume={65},
  number={2},
  pages={920-925},
  doi={10.1109/TAP.2016.2634399}
}

@ARTICLE{MIMO6G,
  author={Wang, Zhe and Zhang, Jiayi and Du, Hongyang and Niyato, Dusit and Cui, Shuguang and Ai, Bo and Debbah, Mérouane and Letaief, Khaled B. and Poor, H. Vincent},
  journal={IEEE Commun. Surveys and Tutorials}, 
  title={A Tutorial on Extremely Large-Scale {MIMO} for 6{G}: Fundamentals, Signal Processing, and Applications}, 
  year={2024},
  volume={26},
  number={3},
  pages={1560-1605},
  doi={10.1109/COMST.2023.3349276}
}

@ARTICLE{FAS,
  author={Wong, Kai-Kit and Shojaeifard, Arman and Tong, Kin-Fai and Zhang, Yangyang},
  journal={IEEE Trans. Wireless Commun}, 
  title={Fluid Antenna Systems}, 
  year={2021},
  volume={20},
  number={3},
  pages={1950-1962},
  doi={10.1109/TWC.2020.3037595}
}

@ARTICLE{PAintro,
  author={Zhang, Yujie and Han, Zixiang and Tang, Shiwen and Shen, Shanpu and Chiu, Chi-Yuk and Murch, Ross},
  journal={IEEE Transactions on Antennas and Propagation}, 
  title={A Highly Pattern-Reconfigurable Planar Antenna With 360° Single- and Multi-Beam Steering}, 
  year={2022},
  volume={70},
  number={8},
  pages={6490-6504},
  doi={10.1109/TAP.2022.3161514}
}

@ARTICLE{PAintro2,
  author={Zhang, Yujie and Tang, Shiwen and Han, Zixiang and Rao, Junhui and Shen, Shanpu and Li, Min and Chiu, Chi-Yuk and Murch, Ross},
  journal={IEEE Transactions on Antennas and Propagation}, 
  title={A Low-Profile Microstrip Vertically Polarized Endfire Antenna With 360° Beam-Scanning and High Beam-Shaping Capability}, 
  year={2022},
  volume={70},
  number={9},
  pages={7691-7702},
  doi={10.1109/TAP.2022.3171342}
}

@article{PIML,
  title     = {Physics-informed machine learning},
  author    = {Karniadakis, George Em and Kevrekidis, Ioannis G. and Lu, Lu and Perdikaris, Paris and Wang, Sifan and Yang, Liu},
  journal   = {Nature Reviews Physics},
  volume    = {3},
  number    = {6},
  pages     = {422--440},
  year      = {2021}
}

@ARTICLE{PAprototype,
  author={Lotfi, Parisa and Soltani, Saber and Murch, Ross D.},
  journal={IEEE Transactions on Antennas and Propagation}, 
  title={Printed Endfire Beam-Steerable Pixel Antenna}, 
  year={2017},
  volume={65},
  number={8},
  pages={3913-3923},
  doi={10.1109/TAP.2017.2716399}
}

\end{document}